\documentstyle[epsfig]{aipproc}

\begin{document}
\title{Collective Evolution of Hot QCD Matter from the QGP to Freeze-Out}
\author{Adrian Dumitru$^*$\thanks{Speaker at CIPANP2000, 7th Conference on the
Intersections of Particle and Nuclear Physics, May 22-28, 2000,
Quebec City, Canada.} 
and Steffen A.\ Bass$^{\dagger}$}
\address{$^*$Physics Department, Columbia University, 538W 120th Street,
New York, NY 10027\\
$^{\dagger}$Nat.\ Supercond.\ Cycl.\ Lab., Michigan State University,
    East Lansing, MI 48824-1321}

\maketitle

\vspace*{-1cm}
\begin{abstract}
We present results on the evolution of $\langle p_t\rangle$ with energy,
for hadrons emerging from hadronization of a quark-gluon plasma possibly
produced in high-energy heavy-ion collisions. We find that BNL-RHIC
energy corresponds to the previously predicted plateau of $\langle p_t\rangle$,
reminiscent of the assumed first-order QCD phase transition.
Heavy hadrons are the best messengers of the transverse flow stall. 
\end{abstract}

Theoretical arguments~\cite{adum:QGP} and lattice QCD
studies~\cite{adum:lattice} suggest that at high temperatures,
roughly $T\sim200$~MeV, the thermodynamically stable state of QCD is different
from that at $T=0$, i.e.\ the theory of strong interactions exhibits a
phase transition to a new state called ``Quark Gluon Plasma'' (QGP).
It is the only phase transition of a fundamental theory accessible to
experiments under defined laboratory conditions. In the QGP the effective
number of relativistic degrees of freedom is expected to be
substantially larger than below the phase transition temperature $T_c$.
The search for signatures of that phase transition is one of the primary
goals of high-energy heavy-ion collision experiments.

In particular, if the transition is first order with non-vanishing
latent heat, it is supposed to leave fingerprints on the hydrodynamical
expansion pattern~\cite{adum:kataja,adum:hung}. Isentropic
expansion would proceed through phase coexistence, where
the pressure $p$ is independent of the energy density $\epsilon$, such that
the isentropic speed of sound $c_s^2={\rm d}p/{\rm d}\epsilon=0$. Thus,
during that phase coexistence stage energy density gradients in the
system do not reflect in pressure gradients, and the average
transverse flow velocity
does not increase substantially. As the $p_t$ spectra of the
emitted hadrons are determined by the transverse boost
velocity and the temperature at freeze-out, one expects
a plateau of $\langle p_t\rangle$
of charged hadrons for some range of beam energies where the central region
``sweeps'' through the mixed phase~\cite{adum:kataja}.

However, the early studies of that effect suffered from some simplifications
which render comparisons to existing experimental data, and extrapolation to
higher energies, unrealistic. First, the latent heat of the transition (and
therefore the space-time volume of the mixed phase) was overestimated
because only pions were assumed to contribute to the entropy of the hadronic
phase. Second, final-state decays of hadronic resonances (like $\rho
\rightarrow\pi\pi$ or $\Delta\rightarrow\pi N$) which contribute more than
50\% to the measured pion multiplicity were not taken into account.
Finally, the most serious problem arises from the arbitrariness of the
decoupling hypersurface, which can not be determined within hydrodynamics but
affects the results considerably.

To reduce the uncertainties and the number of free parameters one therefore
has to improve on the treatment of the post-hadronization stage by considering
microscopic transport~\cite{adum:dum1,adum:BassDum},
e.g.\ within the relativistic Boltzmann equation with collision kernel
as determined by the known hadronic cross sections.
The expansion and subsequent hadronization of the high-temperature
QGP state can be modelled within relativistic
hydrodynamics, assuming that the system evolves locally through a phase
coexistence. Thus, in the space-time region inbetween the initial-time
(purely space-like) hypersurface and the hadronization hypersurface
(the boundary between mixed phase and hadronic phase, which has both space-like
and time-like parts) we solve the continuity equation for the energy-momentum
tensor,
\begin{equation} \label{adum:eqemt}
\partial_\mu T^{\mu\nu}=0~,~{\rm with}~ 
T^{\mu\nu}=(\epsilon+p)u^\mu u^\nu-pg^{\mu\nu}~.
\end{equation}
In the QGP we assume a gas of quasi-free $u$, $d$, $s$
quarks and gluons; for simplicity, the only contribution from interactions
to the QGP pressure is due to the temperature independent vacuum
pressure (bag pressure), see e.g.~\cite{adum:QGP}.

On the hadronization hypersurface, we switch to the Boltzmann equation,
\begin{equation}
p\cdot\partial f_i(x^\mu,p^\nu) = {\cal C}_i~,
\end{equation}
where $f_i(x^\mu,p^\nu)$ denotes the phase-space distribution function of
species $i$.
The binary collision approximation is used to construct the collision kernel
${\cal C}_i$~\cite{adum:uuck},
which treats almost all hadrons from~\cite{Caso:1998tx}
explicitly. All those hadrons are also taken into account in the
equation of state of the mixed phase entering eq.~(\ref{adum:eqemt}).
The switch from hydrodynamics to
microscopic transport is performed by matching the
energy momentum tensors and conserved currents on the hadronization
hypersurface. The evolution towards freeze-out is thus determined
by the relevant hadronic cross-sections and decay rates, competing with the
local expansion rate $\partial_\mu u^\mu$.
The initial condition for central collisions of Pb/Au nuclei and
CERN-SPS energy is that hydrodynamic flow sets in on
the $\tau_i=1$~fm proper time hypersurface, and the entropy per net baryon is
$s/n_B=45$,
which reproduces a variety of measured final-state hadron multiplicities
as well as the $p_t$ spectra of $\pi$, $K$, $p$, $\Lambda$, $\Xi$, and
$\Omega$, see~\cite{adum:dum1,adum:BassDum}. For the higher BNL-RHIC energy,
$\sqrt{s}=200A$~GeV, the parameters were assumed to be $\tau_i=0.6$~fm and
$s/n_B=205$, which yields ${\rm d}N_{ch}/{\rm d}y\approx800$ at $y=0$.
\begin{figure}[b!] 
\centerline{\epsfig{file=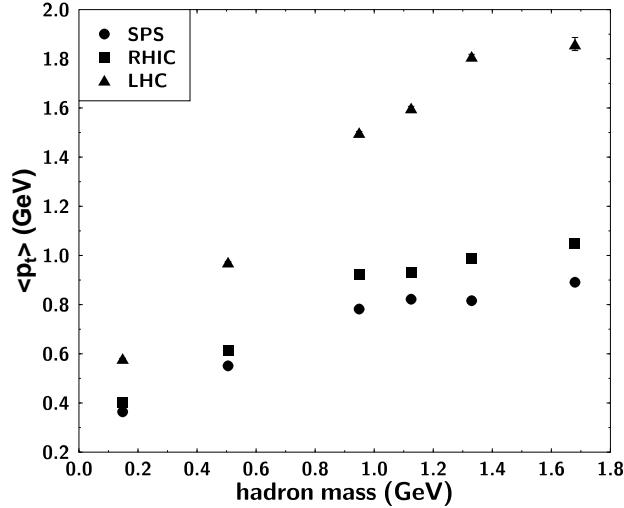,width=3.5in}}
\caption{$\langle p_t\rangle$ versus hadron mass $m$ for 
$\pi$, $K$, $p$, $\Lambda$, $\Xi$, and $\Omega$. Dots for $\sqrt{s}=18A$~GeV,
squares for $\sqrt{s}=200A$~GeV, and triangles for $\sqrt{s}=5500A$~GeV.}
\label{adum:figpt}
\end{figure}
Fig.~\ref{adum:figpt} depicts $\langle p_t\rangle$ for a variety of hadrons,
for CERN-SPS, BNL-RHIC, and CERN-LHC energies. One observes that at each
energy $\langle p_t\rangle$ increases with $m$. Reminiscent of the collective
expansion before decoupling, all hadrons flow approximately with the same
transverse velocity, and therefore their momentum essentially increases in
proportion to their mass. This effect is seen in data obtained for
Pb+Pb collisions at the SPS~\cite{adum:na44}.
The spectra of heavy hadrons are much less affected by the random thermal
motion at decoupling, $v_{th}\propto\sqrt{T/m}$, and are therefore better
suited to measure the transverse flow build up during the evolution.
One also observes that despite the five times higher entropy per baryon
assumed to be achieved at RHIC energy, the average transverse momenta are
predicted by this model to be similar to the lower SPS energy.
This is mainly due to the first-order phase transition featuring a region
of energy densities where $c_s^2$ is small (it is not exactly zero in this
calculation due to the finite conserved baryon charge); and to some
extent also due to earlier decoupling of the hadrons emerging from the
hadronization of the plasma at RHIC energy~\cite{adum:dum1}.

If the initial energy density in the central region continues to grow
towards higher energy as predicted by models assuming saturation of the
transverse density of gluons~\cite{adum:sat} (see however~\cite{adum:dg}),
the hydrodynamical expansion will lead to very large $\langle p_t\rangle$
at LHC energy, see Fig.~\ref{adum:figpt}. Thus, RHIC-energy
is possibly right in the plateau of $\langle p_t\rangle$, if the concept of a
phase mixture is applicable to high-energy collisions.

\begin{figure}[b!] 
\centerline{\epsfig{file=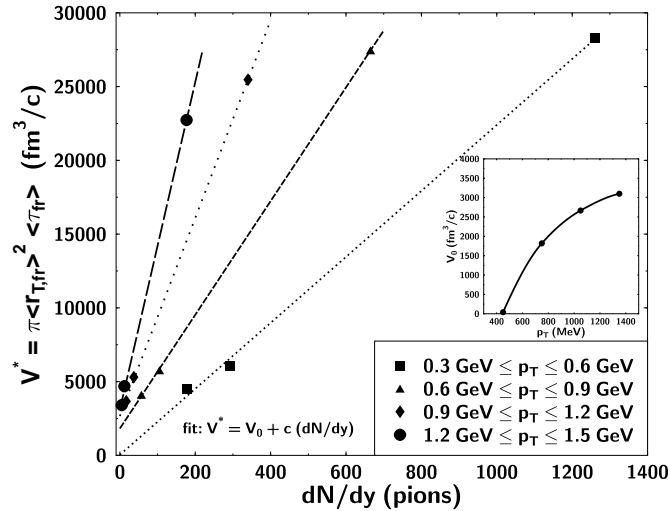,width=3.5in}}
\caption{Freeze-out volume of the pions as a function of the pion
rapidity density at central rapidity, for various $p_t$ cuts.
Increasingly high $p_t$ pions are
only emitted from a ``shell'', the radius of the hollow core increasing
with $p_t$, as shown in the inset.}
\label{adum:figvol}
\end{figure}
Fig.~\ref{adum:figvol} shows the freeze-out volume of the pions at central
rapidity, calculated as the average transverse area $\pi\langle r_t\rangle^2$
times the average length of the central rapidity slice, which is equal to
proper time $\tau$. The space-time volume is essentially increasing linearly
with the multiplicity. Thus, the pion density at freeze-out is not affected
much by the presence of the phase transition, unlike the average transverse
momentum. When extrapolating each line to ${\rm d}N_\pi/{\rm d}y=0$ we obtain
a ``hollow'' volume $V_0$, which increases with $p_t$. This is caused by the
radial flow: high $p_t$ pions can not be emitted from the center,
where the collective velocity field vanishes for symmetry reasons, and
where the pions have only random thermal momenta.

{\bf Acknowledgements:}
A.D.\ thanks the organizers for the invitation to
CIPANP2000 and acknowledges support from a DOE Research Grant, Contract No.\
De-FG-02-93ER-40764. S.A.B.\ is supported by NSF grant PHY-00-70818.


\begin{references}
\bibitem{adum:QGP} e.g.,
Shuryak E., {\it Phys.\ Rept.}\ {\bf 61}, 71 (1980);
McLerran L., {\it Rev.\ Mod.\ Phys.}\ {\bf 58}, 1021 (1986).
\bibitem{adum:lattice} e.g.,
Oevers M., Karsch F., Laermann E., and Schmidt P.,
{\it Nucl.\ Phys.\ Proc.\ Suppl.}\ {\bf 73}, 465 (1999).
\bibitem{adum:kataja}
Kataja M., Ruuskanen P., McLerran L., von Gersdorff H.,
{\it Phys.\ Rev.\ D}  {\bf 34}, 794 and 2755 (1986).
\bibitem{adum:hung}
Blaizot J., and Ollitrault J.,
{\it Phys.\ Rev.\ D} {\bf 36}, 916 (1987);
Hung C., and Shuryak E., {\it Phys.\ Rev.\ Lett.}\ {\bf 75}, 4003 (1995);
Rischke D., and Gyulassy M., {\it Nucl.\ Phys.}\  {\bf A608}, 479 (1996);
Brachmann J., et al., nucl-th/9908010; nucl-th/9912014.
\bibitem{adum:dum1}
Dumitru A., Bass S., Bleicher M., St\"ocker H., and Greiner W.,
{\it Phys.\ Lett.}\  {\bf B460}, 411 (1999).
\bibitem{adum:BassDum}
Bass S., and Dumitru A., {\it Phys.\ Rev.\ C}  {\bf 61}, 064909 (2000).
\bibitem{adum:uuck} Bass S., et al., {\it Prog.\ Part.\ Nucl.\ Phys.}\
{\bf 41}, 225 (1998).
\bibitem{Caso:1998tx}
Caso C., et al., {\it Eur.\ Phys.\ J.}\  {\bf C3}, 1 (1998).
\bibitem{adum:na44}
Bearden I., et al., [NA44 Collaboration],
{\it Phys.\ Rev.\ Lett.}\  {\bf 78}, 2080 (1997).
\bibitem{adum:sat}
Krasnitz A., and Venugopalan R.,
{\it Phys.\ Rev.\ Lett.}\  {\bf 84}, 4309 (2000);
Eskola K., Kajantie K., Ruuskanen P., and Tuominen K.,
{\it Nucl.\ Phys.}\ {\bf B570}, 379 (2000).
\bibitem{adum:dg} 
Dumitru A., and Gyulassy M., hep-ph/0006257.
\end{references}
\end{document}